\documentclass{article}
\usepackage{graphicx}
\usepackage{enumitem}
\PassOptionsToPackage{numbers,sort&compress}{natbib}

\usepackage[dblblindworkshop, final]{neurips_2026}
\workshoptitle{Agentic AI for Biological Discovery (AgenticLS)}

\usepackage[utf8]{inputenc} 
\usepackage[T1]{fontenc}    
\usepackage{hyperref}       
\usepackage{url}            
\usepackage{booktabs}       
\usepackage{amsfonts}       
\usepackage{nicefrac}       
\usepackage{microtype}      
\usepackage{xcolor}         

\title{TIAgent: From Natural Language to Grounded Agentic Workflows for Computational Pathology}

\author{
Aakash Madhav Rao$^{1}$, \quad
Mark Eastwood$^{1}$, \quad
Danielle Belgrave$^{2}$, \quad
Fayyaz Minhas$^{1,}$\thanks{Corresponding Author} \\[1ex]
$^{1}$Predictive Systems in Biomedicine (PRISM) Lab, TIA Centre, Department of \\ Computer Science, University of Warwick, United Kingdom \\
$^{3}$GSK, London, United Kingdom
}

\makeatletter
\providecommand{\bstctlcite}[1]{%
  \@bsphack
  \@for\@citeb:=#1\do{%
    \edef\@citeb{\expandafter\@firstofone\@citeb}%
    \if@filesw\immediate\write\@auxout{\string\citation{\@citeb}}\fi}%
  \@esphack}
\makeatother

\begin{document}
\bstctlcite{BSTcontrol}

\maketitle

\begin{abstract}
  Computational pathology (CPath) analyses increasingly depend on multi-step computational workflows that require substantial programming expertise, limiting their accessibility to pathologists and other domain experts without extensive coding experience. Although large language models (LLMs) provide a promising natural-language interface for constructing such analyses, hallucinated operations and incorrect workflow specifications limit their reliability in scientific settings. We introduce Tissue Image Analysis Agent (TIAgent), an automated CPath workflow-generation framework that, given a natural-language prompt specifying a complex analytical task together with an input dataset, automatically constructs a verifiable computational graph employing existing CPath operations. The resulting graph can be inspected and executed, enabling natural-language construction of complex CPath workflows for applications including automated biomarker discovery and tissue analysis. We demonstrate end-to-end workflow generation and execution and systematically benchmark 14 backbone LLMs across 90 unit tasks spanning 18 CPath operation types. The ability of each LLM to generate a complete and correct unit workflow is quantified using Unit Workflow Success (UWS). Performance varied substantially across backbone models, with Kimi-K2.7-Code achieving the highest UWS of 0.967 (95\% CI: 0.907--0.989). Analysis of node and parameter correctness further revealed that selecting the appropriate computational operation and correctly configuring it represent distinct challenges for LLM-based workflow generation. TIAgent provides a foundation for making complex CPath workflow construction more accessible while enabling systematic evaluation of LLM-driven scientific workflow generation.
\end{abstract}

\section{Introduction}



Computational pathology (CPath) aims to apply computational methods to better understand the underlying patterns within microscopy whole-slide images (WSI). These methods provide quantitative insights into complex tissue characteristics at both individual-image and cohort levels that may not be readily apparent through conventional visual assessment alone. CPath workflows consist of multiple complex analytical steps involving loading images/cohorts, processing, model inference, and post-hoc analysis. Advances in machine learning have further expanded the capabilities of CPath, enabling increasingly sophisticated and diverse approaches to tissue analysis \cite{cpath}. Existing CPath toolkits support a wide range of frequently used functionalities such as WSI processing, visualization, annotation, and inspection of model-derived outputs \cite{Slideflow,TIAToolbox,TIAViz,HistomicsTK}. However, these methods rely on code-based workflows that require significant computational expertise to build and execute, posing a considerable barrier for pathologists and other domain experts with limited programming expertise. 

Large language models (LLMs) offer a potential natural-language interface for automating this process and demonstrate substantial code-generation capabilities \cite{BigCodeBench,xCodeEval,NaturalCodeBench,DA-Code,HumanEval}. However, the risk of hallucinations limits the reliability and verifiability of LLM-generated code for complex workflows. Moreover, performance has been shown to decline on underspecified and compositionally complex tasks, with executable code not necessarily being functionally correct \cite{BigCodeBench,xCodeEval,DA-Code,HumanEval}. Generated programs may contain incorrect parameters, omit required operations, or make unsupported assumptions about the execution environment. This presents a key challenge for agentic workflow authoring: retaining the flexibility of natural-language interaction while grounding generated workflows in valid, executable CPath operations.

To address this challenge, we introduce Tissue Image Analysis Agent (TIAgent), an automated workflow-generation engine that constructs verifiable computational graphs from natural-language scientific requests and input data using existing CPath operations. Our main contributions are:
\begin{itemize}[leftmargin=1em]
\item We introduce, to our knowledge, the first automated workflow-generation engine for constructing CPath workflows through natural-language interaction.
\item TIAgent grounds generated workflows in existing CPath operations and models, constraining the introduction of hallucinated components and enabling computational graphs to be inspected and validated prior to execution.
\item We systematically evaluate TIAgent across 14 backbone LLMs and 90 unit-level tasks spanning 18 CPath operation types, assessing their ability to correctly select and configure the computational components required for workflow construction.
\end{itemize}

\section{Methods}
\begin{figure}[t]
    \centering
    \includegraphics[width=\linewidth]{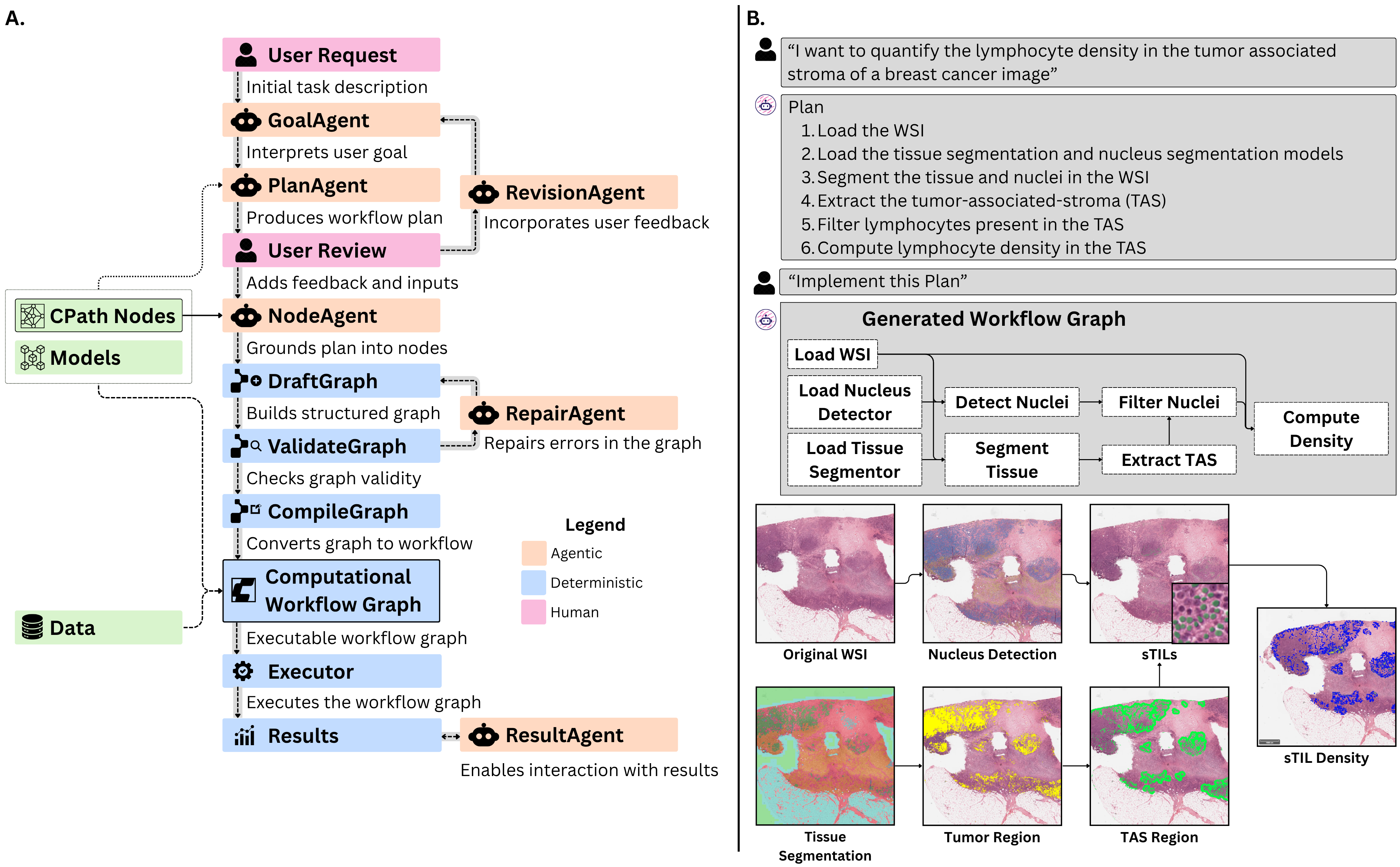}
    \caption{\textbf{A. TIAgent System Architecture:} Given a natural-language request and an input cohort or dataset, TIAgent plans the analysis, selects existing CPath operations and models, and automatically constructs a verifiable computational graph for execution. \textbf{B. End-to-end workflow example:} TIAgent generates and executes a workflow for quantifying stromal tumor-infiltrating lymphocytes (sTILs) within tumor-associated stroma (TAS) in a breast cancer whole-slide image, showing the generated plan and computational graph together with intermediate segmentation and detection outputs and the resulting sTIL density map.}
    \label{fig:architecturediag}
\end{figure}


TIAgent constructs an executable computational graph from a natural-language scientific request and input data (Fig. \ref{fig:architecturediag}A). Given a user request, an LLM-based multi-agent framework first interprets the analytical goal and decomposes it into a structured workflow plan. A \textit{Goal Agent} interprets the request and identifies the intended analytical objective, while a \textit{Plan Agent} decomposes this objective into a sequence of operations and determines the required parameters and models. The resulting plan can be reviewed by the user and modified through a \textit{Revision Agent} before a \textit{Node Agent} grounds each operation to an available CPath node. Subsequent graph construction is deterministic, where the selected nodes, models, parameters, and input data are assembled into a computational graph, validated for structural and parameter-level consistency, and compiled into an executable workflow. Validation failures are returned to a \textit{Repair Agent} for correction before recompilation. The resulting computational workflow graph is passed to an execution engine to produce analytical outputs, which can subsequently be inspected and queried through the agentic interface. This separation between LLM-based planning and node selection and deterministic graph construction, validation, and execution enables natural-language interaction while constraining executable workflows to supported CPath operations.

\paragraph{CPath Nodes and Models} TIAgent constructs workflows from a custom catalogue of CPath nodes and models. CPath nodes represent executable computational operations that can be composed to form a workflow, including data loading, image processing, segmentation, feature extraction, and downstream analysis. Each node defines the operation it performs, its required inputs and parameters, and the outputs it produces, providing a structured interface through which individual operations can be connected within a computational graph. Models include task-specific algorithms that can be supplied as parameters to compatible nodes, allowing a common operation, such as segmentation or feature extraction, to be performed using different models. Separating nodes from models enables the workflow-generation framework to reason independently about which operation is required and, where applicable, which model should perform it. Together, the node and model catalogues define the set of supported CPath capabilities available to the agentic framework, grounding generated workflows in known and executable components.

\paragraph{Benchmark and Evaluation Metrics} We constructed a benchmark of 90 unit-level CPath tasks spanning 18 operation types, with five tasks per operation. Each task specifies a single workflow operation to be generated from a natural-language request and is paired with a manually constructed reference workflow defining the expected node and parameter configuration. We evaluated 14 backbone LLMs, generating one workflow per task and model at a temperature of 0.1, resulting in 1,260 generated workflows. This unit-level evaluation isolates the ability of each backbone model to correctly translate an individual analytical instruction into an executable CPath operation, providing a controlled assessment of the core components required for constructing larger workflows.
Generated workflows were evaluated using three complementary criteria. \textit{Format Validity} ($FV$) measures whether the generated workflow can be successfully constructed and executed without structural or runtime failure. \textit{Node Correctness} ($NC$) measures whether the generated workflow selects the node required by the reference workflow, while \textit{Parameter Correctness} ($PC$) measures whether the selected node is configured with the expected parameter values. We additionally report \textit{Unit Workflow Success} ($UWS$) as a composite measure capturing successful generation of a fully correct unit workflow. Together, these metrics distinguish between generating an executable workflow and correctly selecting and configuring the operation required by the user request.

\section{Results}
\begin{figure}[t]
    \centering
    \includegraphics[width=\linewidth]{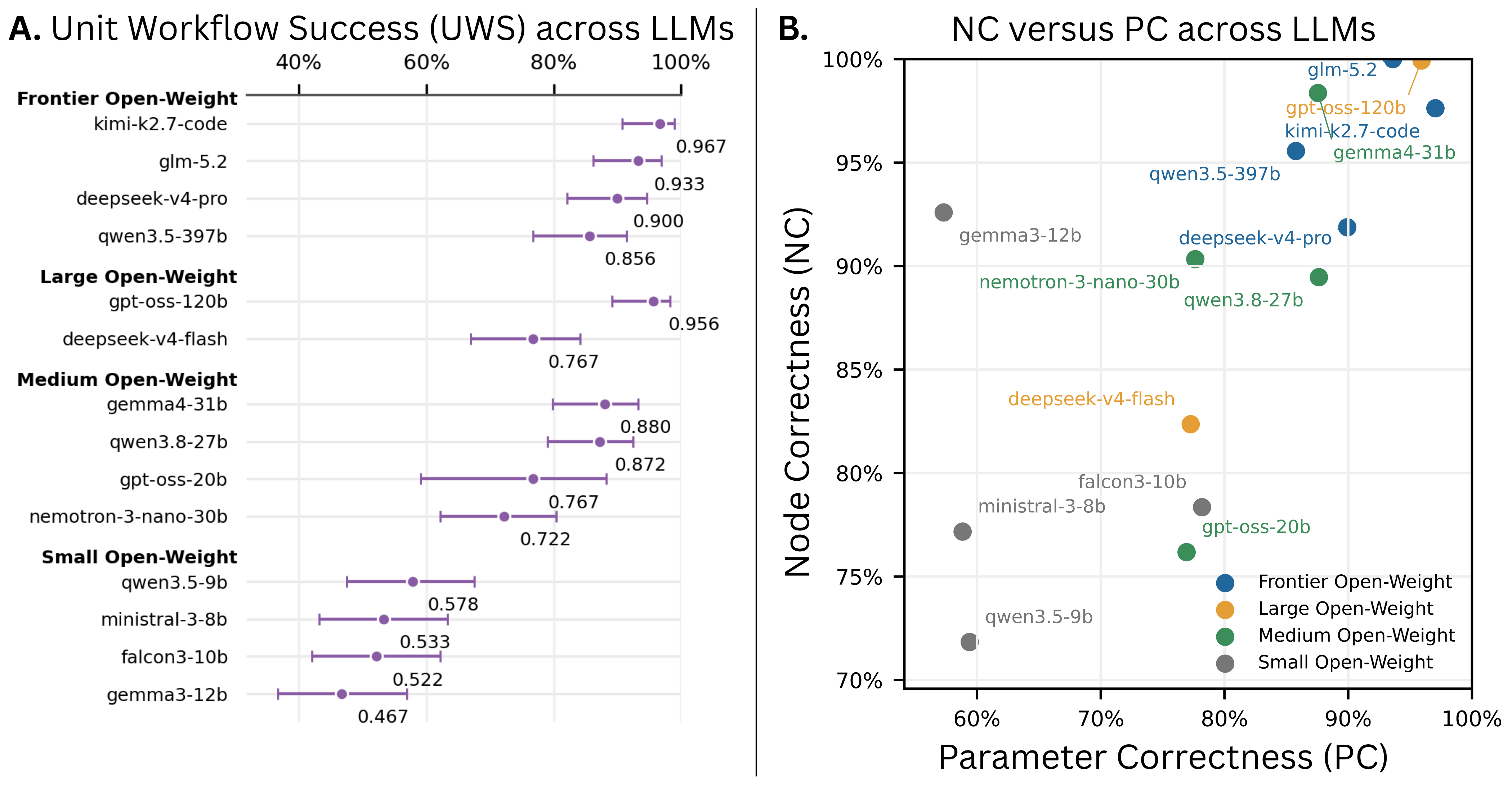}
    \caption{\textbf{A. Unit Workflow Success (UWS) across backbone LLMs.} UWS quantifies each model’s ability to generate a complete and correct executable workflow for individual CPath tasks. \textbf{B. Node and Parameter Correctness across backbone LLMs.} Node Correctness (NC) measures whether the correct computational operation is selected, while Parameter Correctness (PC) measures whether the selected operation is configured with the expected parameters.}
    \label{fig:results}
\end{figure}

To demonstrate end-to-end workflow generation and execution, we applied TIAgent to the quantification of stromal tumor-infiltrating lymphocytes (sTILs) in a breast cancer WSI (Fig. \ref{fig:architecturediag}B). sTILs reflect immune-cell infiltration of tumor-associated stroma and have established prognostic relevance in breast cancer \cite{stils}. Quantifying sTILs computationally requires multiple coordinated image-analysis operations to identify the relevant tissue compartment, detect individual nuclei, distinguish lymphocytes, and determine their distribution within the tumor-associated stroma (TAS). Given the natural-language request, “\textit{I want to quantify the lymphocyte density in the tumor-associated stroma of a breast cancer image}”, TIAgent generated a six-step plan comprising WSI loading, tissue and nucleus segmentation model loading, tissue and nuclei segmentation, TAS extraction, lymphocyte filtering, and lymphocyte-density computation. Following user approval, the plan was grounded into CPath nodes and converted into an executable computational graph connecting the required models and image-processing operations. Execution produced intermediate outputs for tissue segmentation, tumor and TAS identification, and nucleus and lymphocyte detection, culminating in an sTIL density map. This example demonstrates TIAgent's ability to translate a high-level scientific request into a multi-step CPath workflow and execute the resulting computational graph to produce interpretable image-analysis outputs. The complete computational graph generated by TIAgent is shown in Appendix \ref{app:sample-workflow-stils} (Fig. \ref{fig:sampleworkflow}).

Having demonstrated end-to-end workflow generation, we next evaluated the ability of different backbone LLMs to generate the individual CPath operations from which larger workflows are composed. Across the benchmark, UWS varied substantially across the 14 evaluated models (Fig. \ref{fig:results}A), demonstrating the sensitivity of workflow generation to the choice of backbone LLM. To examine the sources of these differences, we separately compared NC and PC (Fig. \ref{fig:results}B). While some models performed similarly across both measures, others showed a marked divergence between their ability to select the correct CPath node and correctly configure its parameters. For example, Gemma3-12B \cite{gemma3} achieved high NC (0.922) but substantially lower PC (0.578), corresponding to a UWS of 0.467. Kimi-K2.7-Code achieved the highest UWS (0.967). These results suggest that identifying the appropriate computational operation and correctly configuring that operation represent distinct challenges for LLM-based workflow generation. Complete results for FV, NC, PC, and UWS, together with their 95\% confidence intervals, are reported in Appendix \ref{app:confidence-intervals} (Table \ref{tab:unit-results-ci}).
\section{Conclusions and Future Work}
We introduced TIAgent, an automated framework that constructs verifiable, executable computational graphs from natural-language scientific requests and input data using existing CPath operations. We demonstrated its ability to generate and execute an end-to-end, multi-step workflow for quantifying stromal tumor-infiltrating lymphocytes in breast cancer and systematically evaluated 14 backbone LLMs across 90 unit-level tasks. Our results reveal substantial differences in unit-level workflow-generation reliability across backbone models, highlighting persistent challenges in translating individual scientific instructions into correctly specified computational operations. While the current benchmark provides a controlled evaluation of the individual operations underlying larger workflows, systematic evaluation of compositionally complex, multi-step workflows remains ongoing. Future work will extend the benchmark to multi-step CPath tasks to assess how reliably these individual capabilities compose into complete analytical workflows.
\small
\bibliographystyle{IEEEtranN}
\bibliography{references}

\newpage
\appendix
\section{Complete Computational Graph for sTIL Quantification}
\label{app:sample-workflow-stils}
\begin{figure}[ht]
    \centering
    \includegraphics[width=\linewidth]{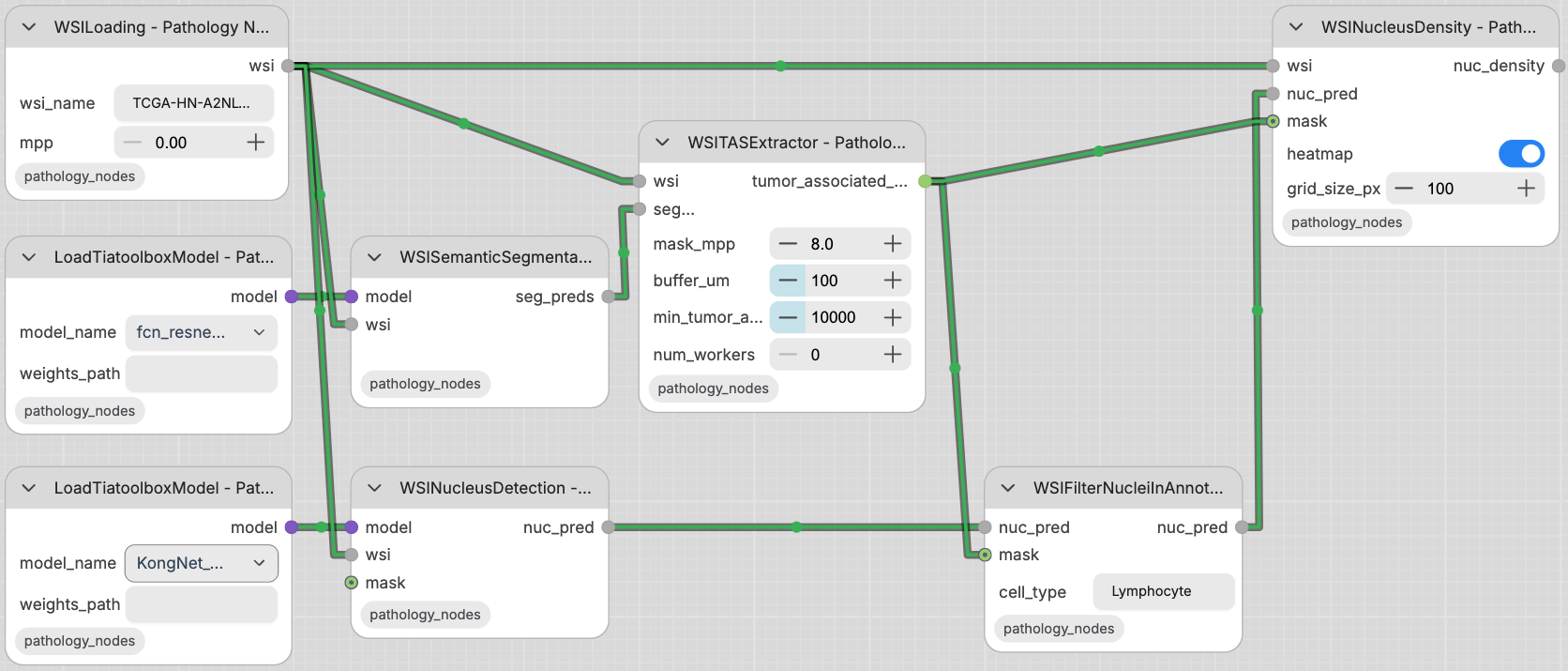}
    \caption{Complete TIAgent-generated computational graph for sTIL density quantification. The workflow loads a breast cancer whole-slide image (WSI) and tissue- and nucleus-analysis models. Tissue segmentation is used to extract the tumor-associated stroma (TAS), while nucleus detection identifies nuclei across the WSI. Detected nuclei are filtered to retain lymphocytes within the TAS, and the resulting predictions are used to compute a spatial sTIL density map. The graph shows the executable node configuration and data flow automatically generated by TIAgent from the natural-language request shown in Fig. \ref{fig:architecturediag}B.}
    \label{fig:sampleworkflow}
\end{figure}

\section{Unit Task Evaluation: Confidence Intervals}
\label{app:confidence-intervals}
\begin{table}[h]
\centering
\caption{TIAgent performance across underlying LLMs on the Unit Task benchmark,
with 95\% Wilson confidence intervals. Each proportion is calculated across
90 unit tasks.}
\label{tab:unit-results-ci}
\small
\resizebox{\textwidth}{!}{%
\begin{tabular}{lcccc}
\toprule
\textbf{LLM} & \textbf{FV (95\% CI)} & \textbf{NC (95\% CI)}
& \textbf{PC (95\% CI)} & \textbf{UWS (95\% CI)} \\
\midrule
GPT-OSS-120B \cite{gptoss}
& 1.000 [0.959, 1.000]
& 1.000 [0.959, 1.000]
& 0.956 [0.891, 0.983]
& 0.956 [0.891, 0.983] \\

DeepSeek-V4-Pro \cite{deepseekv4}
& 0.922 [0.848, 0.962]
& 0.922 [0.848, 0.962]
& 0.900 [0.821, 0.946]
& 0.900 [0.821, 0.946] \\

Qwen3.8-27B \cite{qwen38}
& 0.894 [0.807, 0.939]
& 0.894 [0.807, 0.939]
& 0.872 [0.786, 0.925]
& 0.872 [0.786, 0.925] \\

GPT-OSS-20B \cite{gptoss}
& 0.767 [0.669, 0.842]
& 0.767 [0.669, 0.842]
& 0.767 [0.669, 0.842]
& 0.767 [0.669, 0.842] \\

Gemma3-12B \cite{gemma3}
& 0.789 [0.694, 0.860]
& 0.922 [0.848, 0.962]
& 0.578 [0.475, 0.675]
& 0.467 [0.367, 0.569] \\

DeepSeek-V4-Flash \cite{deepseekv4}
& 0.822 [0.731, 0.888]
& 0.822 [0.731, 0.888]
& 0.767 [0.669, 0.842]
& 0.767 [0.669, 0.842] \\

GLM-5.2 \cite{glm52}
& 1.000 [0.959, 1.000]
& 1.000 [0.959, 1.000]
& 0.933 [0.862, 0.969]
& 0.933 [0.862, 0.969] \\

Kimi-K2.7-Code \cite{kimi}
& 0.978 [0.923, 0.994]
& 0.978 [0.923, 0.994]
& 0.967 [0.907, 0.989]
& 0.967 [0.907, 0.989] \\

Gemma4-31B \cite{gemma4}
& 0.978 [0.923, 0.994]
& 0.978 [0.923, 0.994]
& 0.880 [0.794, 0.930]
& 0.880 [0.794, 0.930] \\

Ministral-3-8B \cite{ministral}
& 0.689 [0.587, 0.775]
& 0.767 [0.669, 0.842]
& 0.589 [0.486, 0.685]
& 0.533 [0.431, 0.633] \\

Nemotron-3-Nano-30B \cite{nemotron3nano}
& 0.856 [0.768, 0.914]
& 0.900 [0.821, 0.946]
& 0.778 [0.682, 0.851]
& 0.722 [0.622, 0.804] \\

Qwen3.5-9B \cite{qwen3.5}
& 0.733 [0.634, 0.814]
& 0.722 [0.622, 0.804]
& 0.589 [0.486, 0.685]
& 0.578 [0.475, 0.675] \\

Qwen3.5-397B \cite{qwen3.5}
& 0.956 [0.891, 0.983]
& 0.956 [0.891, 0.983]
& 0.856 [0.768, 0.914]
& 0.856 [0.768, 0.914] \\

Falcon3-10B \cite{Falcon3}
& 0.678 [0.576, 0.765]
& 0.789 [0.694, 0.860]
& 0.778 [0.682, 0.851]
& 0.522 [0.420, 0.622] \\
\bottomrule
\end{tabular}%
}
\end{table}

\end{document}